\documentclass[aps,prd,twocolumn,superscriptaddress]{revtex4-1} 
\usepackage{fullpage, amsmath, xparse, amssymb, mathtools, graphicx, braket,xcolor,verbatim}
\usepackage{natbib}
\usepackage[colorlinks=true,linkcolor=blue,citecolor=blue]{hyperref}

\definecolor{myblue}{rgb}{ 0.188, 0.478,0.858}

\usepackage{multirow}
\usepackage{hyperref}
\usepackage{cancel}

\begin{document}
\title{Simulations of Shapiro, Gravitational, and Doppler time delays in pulsar networks for ultralight dark matter}
\author{Andrew Eberhardt}
\thanks{Kavli IPMU Fellow. Corresponding author}
\email{\\ andrew.eberhardt@ipmu.jp}
\author{Qiuyue Liang}
\author{Elisa G. M. Ferreira}
\affiliation{Kavli Institute for the Physics and Mathematics of the Universe (WPI), UTIAS, The University of Tokyo, Chiba 277-8583, Japan}
\begin{abstract}
The study of ultralight dark matter helps to constrain the lower bound of the mass in minimally coupled dark matter models. The granular structure of ultralight dark matter density fields produces metric perturbations which have been identified as a potentially interesting probe of this model. For dark matter masses $m \gtrsim 10^{-17} \, \mathrm{eV}$, these perturbations would fluctuate on timescales comparable to observational timescales. In this paper, we estimate the expected time delay these fluctuations would generate in simulated pulsar signals. We simulate arrays of mock pulsars in a fluctuating granular density field. We calculate the expected Shapiro time delay, gravitational redshift, and Doppler shift and compare analytical estimates with the results of simulations. Finally, we provide a comparison with existing pulsar observation sensitivities.
\end{abstract}

\maketitle

\section{Introduction}

Understanding dark matter remains one of the foremost questions in modern cosmology. An international effort combines theoretical, numerical, and observational efforts to search a dark matter parameter space covering over $\sim 80$ orders of decades of mass \cite{Feng2010}. 
At the lowest mass end of this spectrum, typically $m \lesssim 10^{-19} \, \mathrm{eV}$, wave-like phenomena manifest on astrophysical scales and impact large scale structure formation \cite{Hu2000}. 
Moreover, ultraviolet-complete theories, such as string theory, predict a tower of light scalar particles arising from the compactification of extra dimensions \cite{Arvanitaki_2010}, making ultralight dark matter an appealing candidate.

Originally, it was thought that a particle with $m \sim 10^{-22} \, \mathrm{eV}$ could potentially solve the core-cusp \cite{navarro1996, Persic1995, Gentile2004}, missing satellite \cite{Klypin1999, Moore1999}, and too-big-to-fail problems \cite{Boylan-Kolchin2011} which described discrepancies between the predictions of dark matter only simulations and observations of small scale structure (see \cite{Weinberg2015, Bullock2017} for review) without invoking baryonic physics \cite{Hu2000}. 
However, the originally proposed mass of $m \sim 10^{-22} \, \mathrm{eV}$ has since been ruled out by a wide range of phenomena including observations of: 
subhalo mass function ($> 3\times 10^{-21}\, \mathrm{eV}$) \cite{Nadler_2021, Schutz2020},
ultra-faint dwarf half-light radii ($> 3 \times 10^{-19}\, \mathrm{eV}$ \cite{Dalal2022, Marsh:2018zyw}) or cores~\cite{Hayashi:2021xxu}, 
galactic density profiles ($> 10^{-20}\, \mathrm{eV}$ \cite{Bar_2018, Bar_2022}),
satellite masses ($>6 \times 10^{-22}\, \mathrm{eV}$ \cite{Safarzadeh_2020}), 
Lyman-alpha forest ($>2 \times 10^{-20}\, \mathrm{eV}$ \cite{Rogers_2021}), 
strong lensing ($>4 \times 10^{-21}\, \mathrm{eV}$ \cite{Powell:2023jns}). For recent reviews, see \cite{Ferreira_2021,Hui:2021tkt,NIEMEYER2020103787, Eberhardt2025Review}.

Pulsar timing arrays (PTA) are thought to be sensitive to galactic metric perturbations \cite{NANOGrav:2023hvm, IPTA, Khmelnitsky:2013lxt,Porayko:2018sfa,EPTA,Goncharov2021,Verbiest:2024nid}through precise measurements of pulses over $\mathcal{O}(10)$-year observations. This has already been studied in the context of gravitational waves backgrounds \cite{NANOGrav:2023gor, EPTA:2023fyk, Reardon:2023gzh, Xu:2023wog}, substructure of cold dark matter halo \cite{Baghram:2011is,Liu:2023tmv}, primordial black holes \cite{Dror:2019twh}, ultralight dark matter Compton fluctuations \cite{Khmelnitsky:2013lxt,Porayko:2014rfa,Porayko:2018sfa,EuropeanPulsarTimingArray:2023egv}, and ultralight dark matter granule-star interactions \cite{Kim2024}. The effect of higher-spin ultralight dark matter fields on pulsar timing arrays has also been studied \cite{Armaleo:2020yml,Liang:2021bct,Zhang:2023lzt,Wu:2023dnp, Jain2022, Adshead2021, Chowdhury2023}. The gravitational redshift created by ultralight dark matter Compton fluctuations have been used to constrain the dark matter mass around $10^{-23} \, \mathrm{eV}$ \cite{Khmelnitsky:2013lxt,Porayko:2018sfa,EuropeanPulsarTimingArray:2023egv}. Likewise, the impact of ULDM on the frequency of gravitational waves has been studied in \cite{Brax2024, Blas2025}. 

In this paper, we study the Shapiro time delay, gravitational redshift, and Doppler shift caused by de Broglie scale oscillations in the density field. Interestingly, the granular structure oscillates on the time scales probed by experiments, $\sim \mathcal{O}(10) \, \mathrm{yrs}$, for field masses $\gtrsim 10^{-17} \, \mathrm{eV}$. This means that the oscillation of the granular structure may provide some of the highest mass probes of ultralight dark matter. Already de Broglie scale oscillations have been studied in the context of the dynamical heating of galactic stellar populations \cite{Dalal2022, Marsh:2018zyw, DuttaChowdhury2023}, relativistic Doppler shifts in pulsars \cite{Kim2024}, stochastic lensing \cite{Eberhardt2025}, and astrometry \cite{Kim:2024xcr, dror2024}. 

We provide simulations of mock pulsars in an oscillating ultralight dark matter background. We track the Shapiro time delay, gravitational redshift, and Doppler shift between an observer and the pulsars over time and compare simulation results to analytic predictions. We characterize these time delay signals by calculating the root-mean-square amplitude, and dependence on dark matter and pulsar-observer parameters. Finally, we compare the result of this work to the existing ultralight dark-matter constraints from pulsar timing arrays, and estimate the sensitivity of current PTA experiments to these granular phenomena. We conclude that both the Shapiro time delay and gravitational redshift effect of the granules would produce unique temporal power spectra in pulsar timing array systems, which, if observed, would provide a smoking gun for ultralight dark matter. Our simulations of the Doppler shift corroborate the conclusions of \cite{Kim2024}. However, current experiments are unlikely to be sensitive to any of these effects.

The paper is organized as follows. In Section \ref{sec:background} we provide a background on ultralight dark matter and gravitational time delays. Section \ref{sec:simulations} includes a description of simulations. We discuss results in Section \ref{sec:results} and observational implications in Section \ref{sec:observations}. Finally, we conclude in Section \ref{sec:conclusions}.  

\section{Background} \label{sec:background}

\begin{figure}
	\includegraphics[width = .48\textwidth]{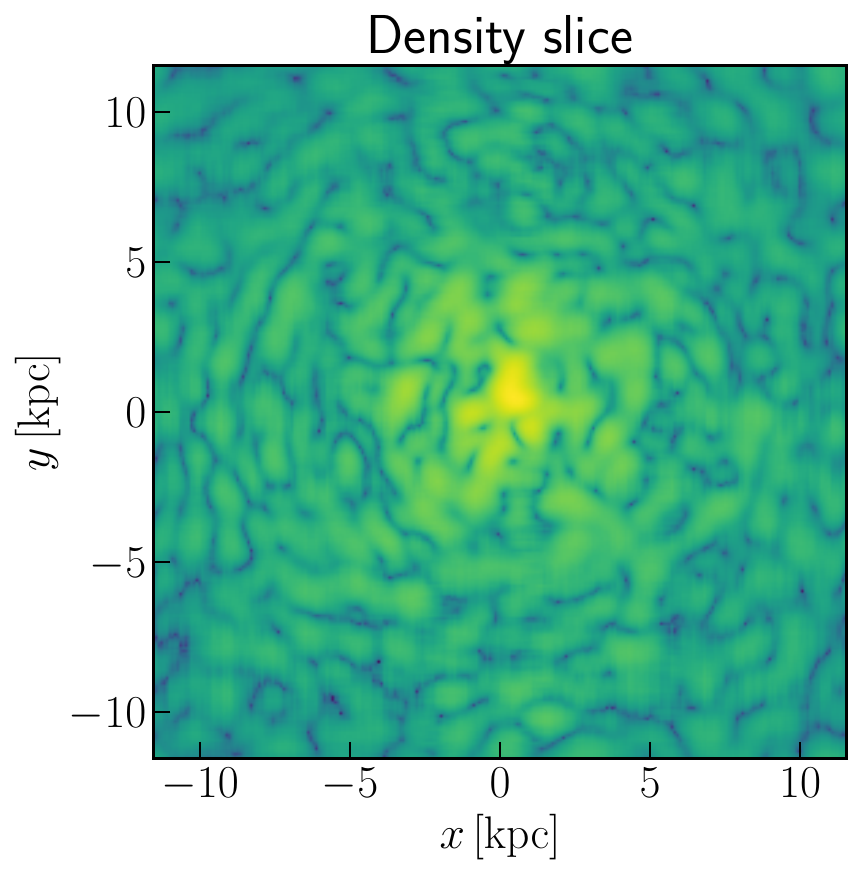}
	\caption{ A density slice through a typical dark matter halo. Here we plot the log density. We can see the granule structure resulting from interference patterns in the classical field. The normalization of the field is arbitrary. }
	\label{fig:rho_halo}
\end{figure}

\subsection{Ultra light dark matter}
Ultralight dark matter is usually treated as a classical \cite{Eberhardt2023} and non-relativistic wavelike field. In this paper, we will assume that it is a scalar field, but generalization of higher spins also exists \cite{Jain2022,Adshead:2021kvl}. We start with the Klein-Gordon equation for a quadratic potential
\begin{align} \label{eqn:KG_eqn}
    \left( \square^2 + \frac{m^2 c^2}{\hbar^2} \right) \phi = 0 \, .
\end{align} 
The Klein-Gordon field can be split into its slowly and quickly oscillating portions by introducing a complex field, $\psi$, and writing the Klein-Gordon field as 
\begin{align}\label{eqn:nonRelApprox}
    \phi = \psi \, e^{-i\,m c^2\,t / \hbar} + \mathrm{c.c} \, .
\end{align} 
The field $\psi$ varies spatially on the de Broglie scale, $\lambda_\mathrm{db}$, and on the de Broglie time, $\tau_\mathrm{db}$, and the exponential carries with it the quickly oscillating portion which oscillates on the Compton timescale, $\tau_c= 2 \pi \hbar / m c^2$. In this work, we will be concerned with the slowly varying portion of the field. To study the de Broglie scales effects we plug equation \eqref{eqn:nonRelApprox} into \eqref{eqn:KG_eqn} and factor out the quickly oscillating terms. Then taking the weak gravity limit, we arrive at the Schr\"odinger-Poisson equations which describe the evolution of a non-relativistic spin 0 field, given
\begin{align}
    \partial_t \psi &= \frac{-i}{\hbar} \left( \frac{-\hbar^2 \nabla^2}{2m} + m \Phi \right) \psi \, , \\
    \nabla^2 \Phi &= 4 \pi G |\psi|^2 \, , \label{eqn:SP}
\end{align} 
and relates the Newtonian potential, $\Phi$ to the field amplitude. Here we have chosen the field norm so that $|\psi|^2 = \rho$ is the spatial density field.  
 
The time scale for the quickly and slowly oscillating parts are the Compton time, $\tau_\mathrm{c}$, and de Broglie time, $\tau_\mathrm{db}$, respectively. The slowly varying part also varies spatially on the scale of the de Broglie wavelength, $\lambda_\mathrm{db}$. These quantities are estimated as functions of the mass,
\begin{widetext}
\begin{align}
    \tau_\mathrm{c} &= 2 \pi \hbar / m c^2 \sim 10^{-5} \left( \frac{10^{-17} \, \mathrm{eV}}{m} \right) \, \mathrm{yrs}  \, , \\
    \tau_\mathrm{db} &= 2 \pi \hbar / m \sigma^2 \sim 30 \left( \frac{10^{-17} \, \mathrm{eV}}{m} \right) \left( \frac{200 \, \mathrm{km/s}}{\sigma} \right)^{2} \, \mathrm{yrs} \, ,  \\ 
    \lambda_\mathrm{db} &= 2 \pi \hbar / m \sigma \sim 6 \times 10^{-6} \left( \frac{10^{-17} \, \mathrm{eV}}{m} \right)  \left( \frac{200 \, \mathrm{km/s}}{\sigma} \right) \, \mathrm{kpc} \, .
\end{align}
\end{widetext}
Where in this work $\sigma$ is the velocity dispersion of the dark matter halo. In the following work, $\sigma$ will generally refer to the the velocity dispersion of the system. We note here that the definition of these time and length scales often differs in the literature up to a factor of $2 \pi$, for example our notation differs from \cite{Kim2024}. However, keeping definitions consistent, we will recover similar results.

The spatial interference between velocity streams gives rise to a granular interference pattern in dark matter halos, see Figure \ref{fig:rho_halo}. It has been found that the distribution of overdensities around the radial averaged profile, i.e. $\delta \rho(\vec r) = \rho(\vec r) - \braket{\rho(r)}$, is well described by the overdensities arising from the superposition of plane waves with a velocity dispersion given by a Maxwell-Boltzmann distribution \cite{Dalal_2021}, where $\rho(\vec r)$ is the density at the point $\vec r$ in our halo and $\braket{\rho(r)}$ is the average value of the density in the halo at $r = |\vec r|$. Here we assume that the halo is isotropic.

Therefore, if we study the overdensities in the vicinity of some point, $\vec R$, in a halo, they are well described by the following approximation for the field
\begin{align}
    \psi(\vec r) &= \sqrt{\braket{\rho(|\vec R + \vec r|)}} \int d^3 \vec v \sqrt{f(v)} \, e^{- i m \vec v \cdot \vec r / \hbar} \, , \\
    f(v ) &= \sqrt{\frac{2}{\pi}} \frac{v^2}{\sigma^2} e^{-v^2 / 2 \sigma^2} \, ,
\end{align}
where $\sigma$ is the dark matter velocity dispersion at $\vec R$. This holds as long as we only look over scales that are smaller than the galactic halo scale radius, i.e. $r \lesssim R_s$. We will take the typical Earth-pulsar separation to be $\sim 1 \, \mathrm{kpc}$, which is smaller than the scale radius of the Milky-way halo ($R_s \sim 2 \, \mathrm{kpc}$ \cite{porcel1997}). 

The interference patterns create granule structures in the density. It is often a useful approximation to treat these granules as quasiparticles with an effective radius approximately equal to the de Broglie wavelength $r_\mathrm{eff} \sim \lambda_\mathrm{db}$, and effective mass $m_\mathrm{eff} \sim \lambda_\mathrm{db}^3 \rho$ \cite{Dalal2022, Kim2024}. The granules persist for approximately the crossing time of the granule which is given by the de Broglie time $\tau_\mathrm{db}$.

\subsection{Gravitational time delays}

In the weak gravity limit appropriate for describing ultra light dark matter we can write the metric as 
\begin{align}
    ds^2 = \left(1 + 2 \frac{\Phi}{c^2}\right) \, c^2 dt^2 - \left(1-2\frac{\Psi}{c^2} \right) \, dx^2 \, ,
\end{align}
where we will also assume that $\Phi = \Psi$ is just the Newtonian potential calculated in \eqref{eqn:SP}. 

\subsubsection{Redshift and Shapiro delay}
This potential then creates a redshift, at time $t$, in a pulsar with frequency $\Omega_p$ at position $r_p$, as observed by an observer at $r_e$ given
\begin{align} \label{eqn:frequencyShift}
    z(t) \equiv \frac{\Delta \Omega_p(t)}{\Omega_p} &= \frac{\Phi(\vec r_e, t) - \Phi(\vec r_p, t')}{c^2} \\
    &- \int_{t'}^t 2 \hat n_i \, \partial_i \frac{\Phi(\vec x,t'')}{c^2} \, dt'' \, , \nonumber
\end{align}
where $\hat n_i = (\vec r_e - \vec r_p) / |(\vec r_e - \vec r_p)|$ is the unit vector  in the direction of signal propagation and $x(t'')$ is the path of the signal from the pulsar to observer. $t$ is the time the pulse is observed at Earth, and $t'$ is the time it is emitted from the pulsar. The first line in equation \eqref{eqn:frequencyShift} gives the gravitational redshift of the frequency, and the second the Shapiro time delay shift in the frequency. The effects then given a time delay via integration
\begin{align}
    \delta t(T) = \int_0^T \frac{\Delta \Omega_p(t')}{\Omega_p} dt' \, ,
\end{align}
where $T$ is the total observation time.

\subsubsection{Doppler shift}

The fluctuating potential gradients also impact the proper motion between observer and pulsar. This relative proper motion results in a Doppler shift of the observed frequency of the pulsar. This shift can be found by integrating the relative forces due to the gravitational potential, $\Phi$, giving a frequency shift due to the Doppler effect of
\begin{align}
    z_D = \int^t \hat n_i \partial_i \left( \frac{\Phi(t' - D/c, \, r_p) - \Phi(t', \, r_e)}{c} \right) \, dt' .
\end{align}
Where $D$ is the distance between the pulsar and the observer, approximated as constant over the observing time $t$. 

\section{Simulations} \label{sec:simulations}

This section describes the simulations used in this work. The code used to produce these simulations and the figures in this draft are publicly available in \href{https://github.com/andillio/PTA_FDM_public}{andillio/PTA\_FDM\_public}. Data used in this paper can be made available upon request. 

\begin{figure*}[!ht]
	\includegraphics[width = .97\textwidth]{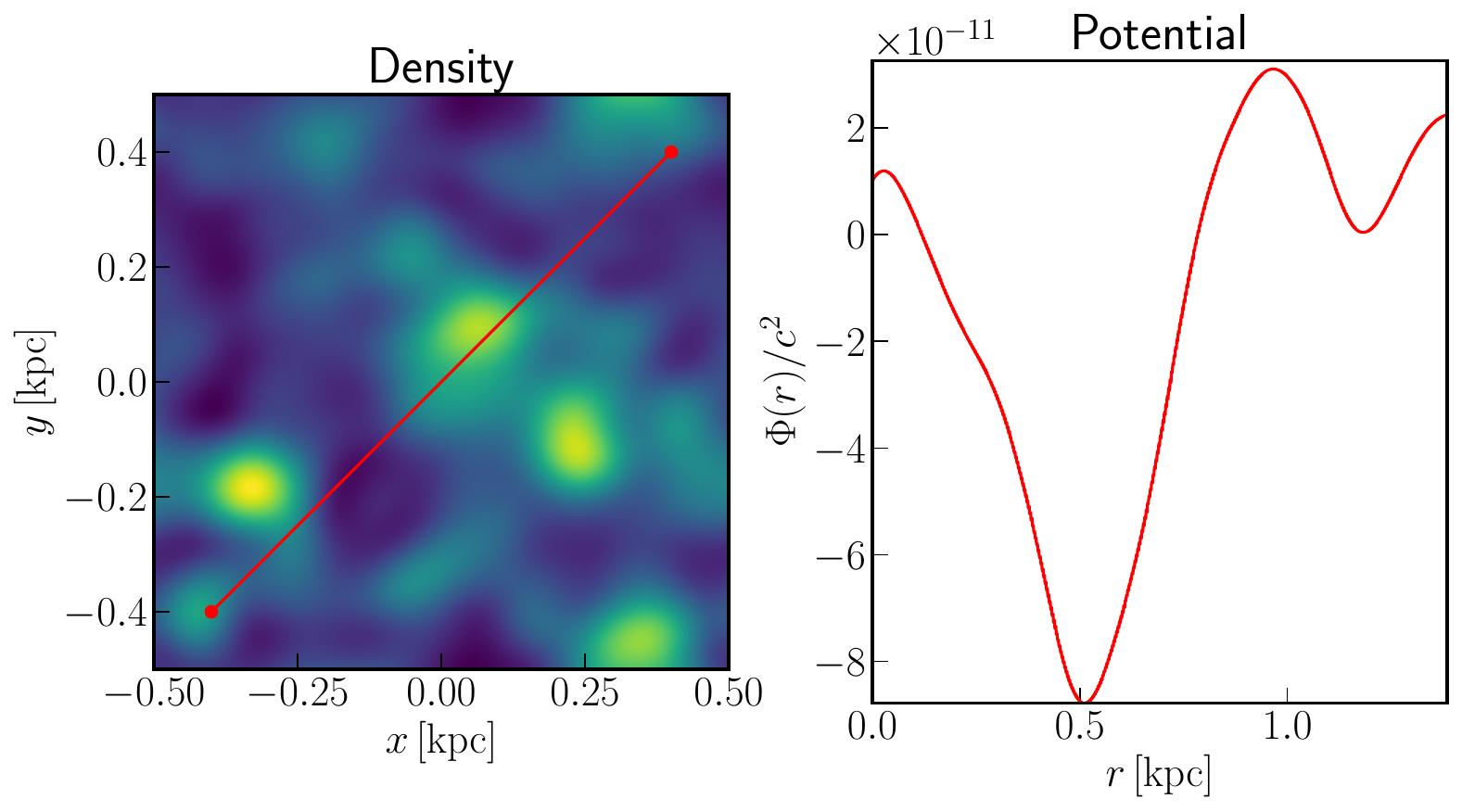}
	\caption{ The projected spatial ultralight dark matter density is shown on the left. The superimposed line in red on the left plot is a projected hypothetical path between a pulsar and Earth. The value of the potential interpolated at each position is potted on the right. Here the field mass $m = 10^{-22} \, \mathrm{eV}$.}
	\label{fig:simDensity}
\end{figure*}

\subsection{Initial conditions} \label{sec:ICs}

We initialize the field as a sum of plane waves in a box of length $L$ with periodic boundary conditions. The field is given at position, $\vec r$, as 
\begin{align} \label{eqn:planeWaveSum}
    \psi(\vec r) = \sqrt{\rho_0} \sum_i^{n_s} e^{- i m \vec v_i \cdot \vec r / \hbar} \, ,
\end{align}
where $\rho_0 = 10^7 \, \mathrm{M_\odot / kpc^3}$ is the average density of the system. $n_s$ is the number of different velocity streams used in the construction of the field. It is important that this is $n_s \gg 1$, we typically use $n_s \sim 2000$. $\vec v_i$ is the velocity of the $i$th plane wave, chosen isotropically and with magnitude drawn from a Maxwell-Boltzmann distribution
\begin{align}
    f(v) = \sqrt{\frac{2}{\pi}} \frac{v^2}{\sigma^2} e^{-v^2 / 2 \sigma^2} \, , \label{eqn:maxwellian}
\end{align}
$\sigma$ is the velocity dispersion of the dark matter. We will take $\sigma = 200 \, \mathrm{km/s}$. The resulting density pattern can be seen in the left panel of Figure \ref{fig:simDensity}. The power spectrum of fluctuations created by these initial conditions consistently models the fluctuations around the average density profile in typical ultralight dark matter halos \cite{Hui_2021_vortices,Dalal_2021}, see Appendix \ref{sec:appendix_density_ps}.
\subsection{Solver}

The equations of motion are integrated only using the kinetic term in the Hamiltonian, i.e. dynamical gravitational effects are not included. The field is updated from time $t$ to time $t+\Delta t$ as follows
\begin{align}
    &\psi(t + \Delta t) = \mathcal{F}^{-1}\left[ e^{-i\hbar k^2 \Delta t / 2 m} 
    \mathcal{F}\left[ \psi_i(t) \right] \right] \, ,
\end{align}
where $\mathcal{F}$ is a discrete fast Fourier transform. We note that because the system does not evolve under its own gravitational potential that the time step $\Delta t$ can be chosen arbitrarily. In principle, we could integrate equation \eqref{eqn:planeWaveSum} directly and then reconstruct the field at any time by re-summing these plane-waves. This method would have a similar runtime depending on the simulation parameters chosen.

\subsection{Gravitational redshift and Shapiro delay}
\begin{figure*}[!ht]
	\includegraphics[width = .97\textwidth]{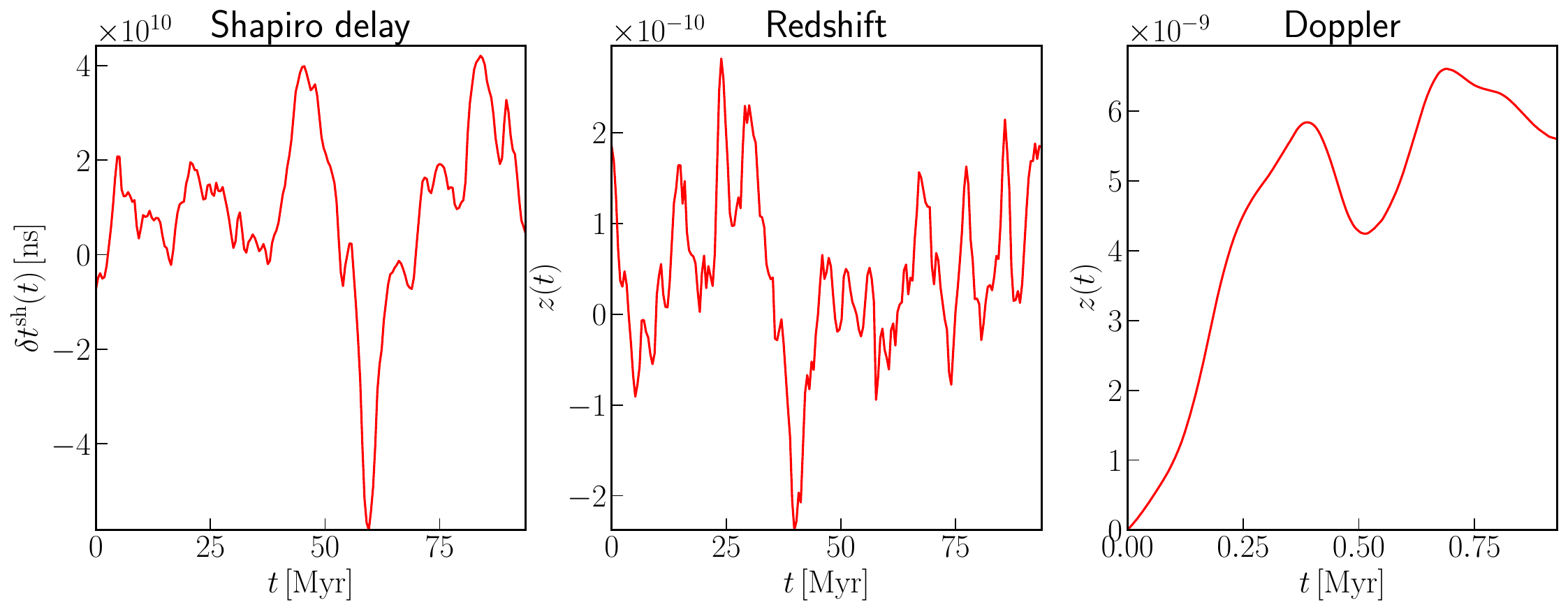}
	\caption{ The Shapiro time delay (left), gravitational redshift (center), and Doppler shift (right) for a simulated pulsar at $1 \, \mathrm{kpc}$ from Earth in a ULDM density field at the local dark matter density. The mass of the dark matter field in this simulation was $m = 10^{-22} \, \mathrm{eV}$. }
	\label{fig:signal}
\end{figure*}
Pulsar and observer locations are chosen within the simulation volume, $r_p$ and $r_e$ respectively. The Shapiro time delay and gravitational redshift between each pulsar and Earth are calculated, at each timestep, $t_i$, see Figure \ref{fig:signal}. The expected root-mean-square Shapiro time delay between Earth and the $j$th pulsar at time $t_i$ can be estimated
\begin{align}
    \braket{\delta t_j^{sh}(t_i)}_\mathrm{rms} = \int_{r^j_p}^{r_e} \frac{dl}{c} \, \frac{\delta \Phi(l, t_i)}{c^2} \, ,
\end{align}
where $r^j_p$ is the location of the $j$th pulsar. This approximation is discussed in more detail in Appendix \ref{app:approxShapiro}. The integrand is plotted for an example simulation realization in the right panel of Figure \ref{fig:simDensity}.

Notice that this approximates the potential as stationary over the travel time of the pulse from the pulsar to Earth. This is not the case in general, however, the statistics of the potential remain constant over the travel time. In principle, this approximation could introduce specious spatial correlations. But, as long as the travel distance of the pulse is large compared to the spatial correlation length, $|\vec r_p -\vec r_e| \gg \hbar / m \sigma$, this approximation should be accurate. Note also, we do not account for the relative motion between the Earth and pulsar, the effect this has on the Doppler shift of the pulsar frequency is studied in \cite{Kim2024}.

The gravitational redshift for the $j$th pulsar at time $t_i$ is given
\begin{align}
    z_j(t_i) = \frac{\Phi ( \vec r_e) - \Phi(\vec r_p^j)}{c^2} \,, \label{eqn:num_z}
\end{align}
$\Phi$ is the solution to Poisson's equation for the classical field, 
\begin{align}
    \nabla^2 \Phi(\vec r) = 4 \pi G \rho(\vec r) \, . 
\end{align}

We solve for this potential numerically using the spectral method, i.e. 
\begin{align}
    \Phi(\vec r) = \mathcal{F}^{-1}\left[ 4 \pi G \,  \frac{\mathcal{F} [\rho(\vec r)] (\vec k)}{k^2} \right] \, . \label{eqn:Poisson_k_space}
\end{align}
When performing the cumulative integral of equation \eqref{eqn:num_z} we remove the mean of the data as it would be indistinguishable from the intrinsic pulsar frequency. 

\subsection{Doppler shift}

To model the Doppler shift we need to track the relative motion of pulsars not just the gravitational potential. Therefore, in these simulations we treat the pulsars as an ensemble of massless point particles with uniform random initial positions in the box and velocities drawn from the same Maxwell-Boltzmann distribution as the dark matter, i.e., drawn from distribution in equation \eqref{eqn:maxwellian}. The point particles are coupled to the gravitational potential of the ultralight dark matter and updated using a drift-kick-drift scheme where the drift half-step now involves both an update of the ultralight dark matter field and the $i$th particle as 
\begin{align}
    \psi(t + \Delta t/2) &= \mathcal{F}^{-1}\left[ e^{-i\hbar k^2 \Delta t / 4 m} 
    \mathcal{F}\left[ \psi_i(t) \right] \right] \, , \\
    \vec r_i(t + \Delta t /2) &= \vec r_i + \vec v_i(t) \, \Delta t / 2 \, ,
\end{align}
and a kick full-step as
\begin{align}
    \vec v_i(t + \Delta t) = \vec v_i + \nabla \phi(\vec r_i) \, \Delta t \, .
\end{align}
Where the potential gradient is calculated using a finite (5-pt) differentiation stencil and linearly interpolated at the particles position. 

The Doppler shift for the $i$th particle at each data drop can be calculated using the velocity along the line of sight to the observer as 
\begin{align}
    z_D^i(t) = \frac{\hat n_i \cdot \vec v_i}{c} \, .
\end{align}

\section{Results} \label{sec:results}

\begin{figure*}[!ht]
	\includegraphics[width = .97\textwidth]{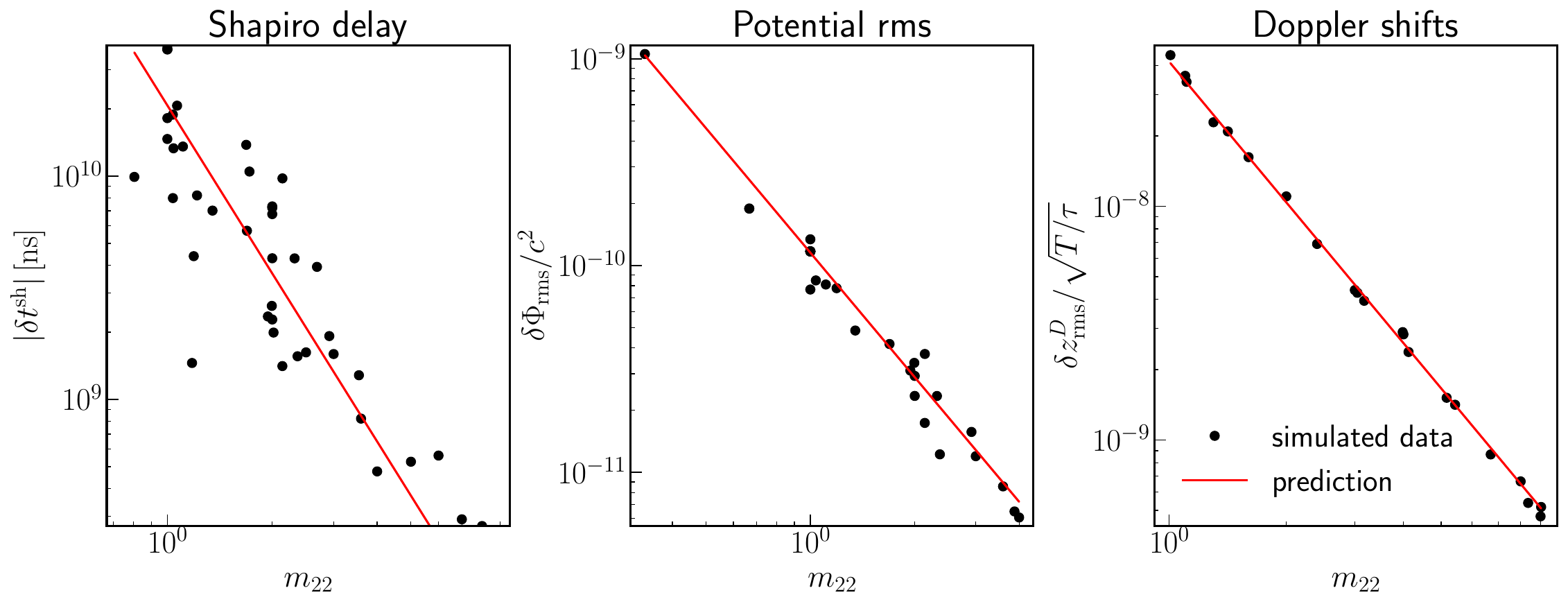}
	\caption{ The average Shapiro time delay measured for pulsars at a distance of $1 \, \mathrm{kpc}$ (left) and the average gravitational potential fluctuations (center), $\braket{\Phi}_\mathrm{rms}$, the average Doppler shift per root time (right), for a set of plane-wave box simulations, described in section \ref{sec:ICs}, as a function of the field mass $m_{22}$. The data points for the Shapiro delays correspond to the average over all the pulsars at a single moment in time. The large scatter around the prediction is due to the random potential fluctuations around the observer at that time in each simulation. The quasi-particle approximation for each effect equation is shown in red. For Shapiro delays the prediction is equation \eqref{eqn:dt_sh}, for the potential rms the prediction is given by equation \eqref{eqn:phi_rms_approx}, and for the Doppler shifts the prediction is given by equation \eqref{eqn:doppler_z_over_T}. The quasi-particle approximation provides a reasonably accurate estimation in each situation.  }
	\label{fig:QP_prediction}
\end{figure*}

\begin{figure*}[!ht]
	\includegraphics[width = .97\textwidth]{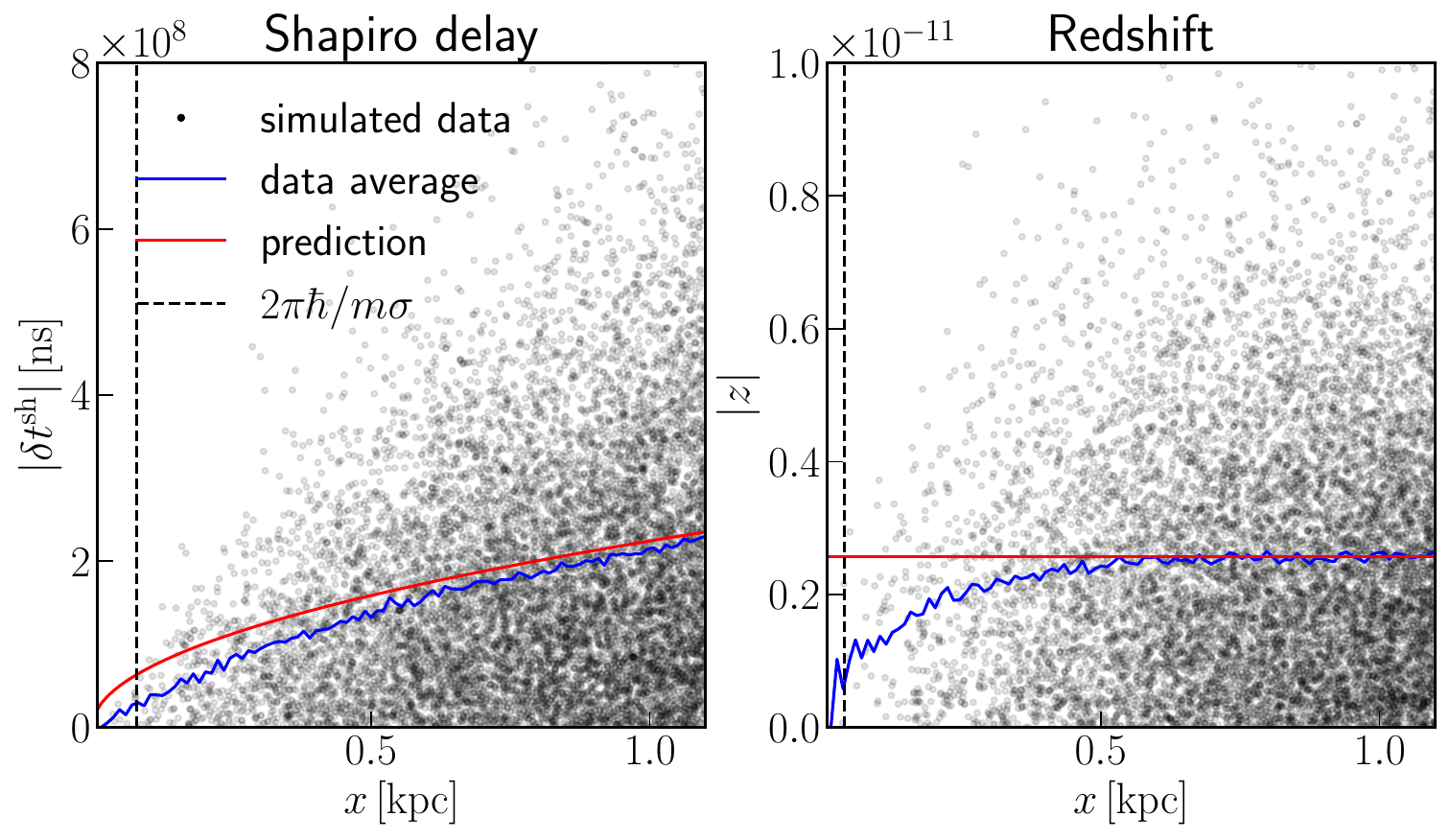}
	\caption{ The average Shapiro time delay (left) and gravitational redshift (right) measured for pulsars in a plane-wave box simulation described in section \ref{sec:ICs} as a function of the pulsar-Earth separation, $x$. The quasi-particle approximation, equation \eqref{eqn:phi_rms_approx}, is shown in red. Time delay values are shown as blue dots for specific pulsar-Earth pairs, with the average as a function of $x$ plotted in blue. The quasi-particle approximation provides a reasonably accurate estimation of the Shapiro time delays in simulations. Note that the amplitude of the Shapiro delay grows with the distance approximately as a random walk. However, the redshift values grow quickly over a few de Broglie wavelengths and then quickly plateau near the predicted value. The field mass for this simulation was $7.5 \, m_{22}$. Pulsar and Earth position pairs were chosen randomly in a box of size $L = 2.2 \, \mathrm{kpc}$. }
	\label{fig:distance}
\end{figure*}
\begin{figure*}[!ht]
	\includegraphics[width = .97\textwidth]{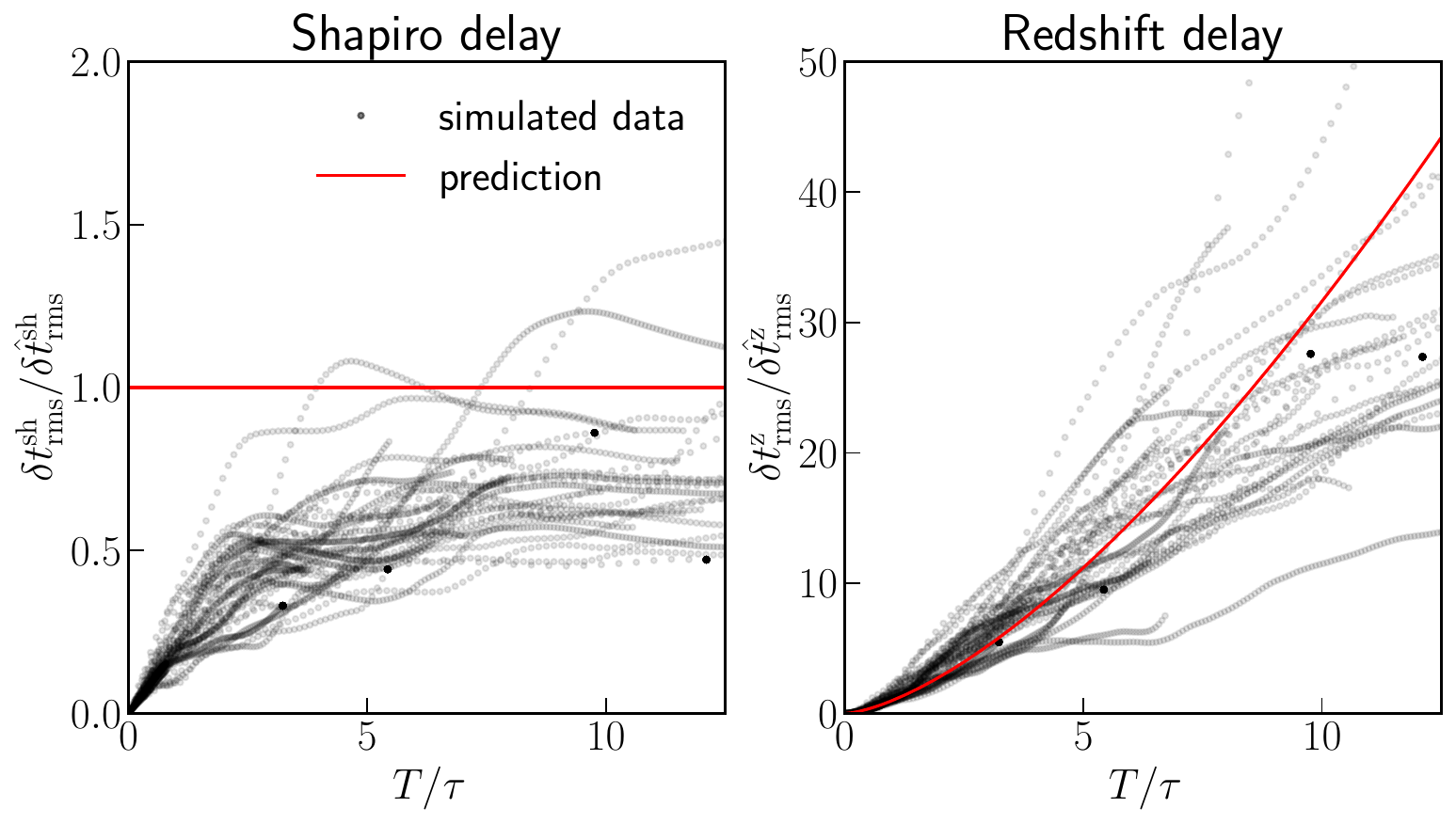}
	\caption{ The root-mean-square time delays as a function of integration time $T$ for a set of simulations at a variety of masses from $10^{-22} \, \mathrm{eV} < m < 10^{-21} \, \mathrm{eV}$. The left panel shows the time delays associated with the Shapiro time delay, and the right the time delay associated with the gravitational redshift. The time for each simulation is scaled by its de Broglie time and the time delays by the predicted time delay in equations \eqref{eqn:dt_sh} or \eqref{eqn:dt_z}. The red dashed line shows the predicted time dependence of each signal.}
	\label{fig:intTime}
\end{figure*}

Throughout this section, we will approximate quantities using the quasi-pariticle approximation and then compare these approximations to the results of simulated data. In this approximation we assume that the density field can be described by a collection of tightly packed ``quasi-particles" modeled as spheres with radius equal to the de Broglie wavelength, $\lambda_\mathrm{db}$, and effective mass $m_\mathrm{eff} \sim \rho \lambda^3_\mathrm{db}$, with $\rho$ the local field density. In the context of ultralight dark matter, this approximation has been previously applied to the description of the dynamical heating of dwarf galaxies \cite{Dalal2022}, pulsar Doppler shifts \cite{Kim2024}, and astrometry \cite{Kim:2024xcr}.

\subsection{Shapiro time delay}
We can use the quasi-particle approximation to estimate the magnitude of the Shapiro delay. The granules are tightly packed, so we expect to encounter $N_\mathrm{enc} \sim D/\lambda_\mathrm{db}$ granules along the path from the pulsar to the Earth, where $D$ is the Earth pulsar separation, which we will usually take to be $\sim 1 \, \mathrm{kpc}$. The granules can add both positive and negative delays randomly and so we expect the root-mean-square value of the delay to grow approximately as the root of the number of encounters as long as $D/\lambda_\mathrm{db} \gg 1$, which should be the case for masses $m \gg 10^{-22} \, \mathrm{eV}$. We find in simulations that the root-mean-square time delay scales as $N_\mathrm{enc}^{1/2}$ at separations large compared to the de Broglie wavelength, see Figure \ref{fig:distance}.

Each granule gives a time delay of order 
\begin{align}
    \delta t_\mathrm{rms} / \mathrm{granule} &\sim \Phi_\mathrm{eff} / c^2 t_\mathrm{cr} \sim
    \frac{G \, m_\mathrm{eff}}{c^2 \, \lambda_\mathrm{db}} \frac{\lambda_\mathrm{db}}{c} \nonumber \\
    &\sim G m_\mathrm{eff} / c^3
\end{align}
which is just the potential at the de Broglie radius, $\Phi_\mathrm{eff} \sim \frac{G m_\mathrm{eff}}{\lambda_\mathrm{db}}$, multiplied by the granule light crossing time, $t_\mathrm{cr} \sim \lambda_\mathrm{db}/c$. This approximation is discussed in more detail in Appendix \ref{app:approxShapiro}. We can therefore approximate the Shapiro time delay as 
\begin{widetext}
\begin{align}
    &\delta t^{sh}_{rms} = d \, N_\mathrm{enc}^{1/2} \left( dt_\mathrm{rms} / \mathrm{granule} \right) = d \frac{G \, m_\mathrm{eff}}{c^3} \left(\frac{D}{\lambda_\mathrm{db}}\right)^{1/2} \, , \label{eqn:dt_sh} \\
    & = d\, G \, \rho \frac{\hbar^{5/2} D^{1/2}}{m^{5/2} \, \sigma^{5/2} \, c^3} \, , \nonumber \\
    &\sim 7 \times 10^{-3} \left( \frac{10^{-17} \, \mathrm{eV}}{m} \right)^{5/2} \left( \frac{200 \, \mathrm{km/s}}{\sigma} \right)^{5/2} \left( \frac{\rho}{10^7 \, \mathrm{M_\odot / kpc^3}} \right) \left( \frac{D}{\mathrm{kpc}} \right)^{1/2} \, \mathrm{ns}\ , \nonumber 
\end{align}
\end{widetext}
where we find from our simulations that $d \approx 1.5 \sim \mathcal{O}(1)$ is some $\mathcal{O}(1)$ constant. This approximation provides a reasonably accurate estimate of the time delays measured in our simulations, see figures \ref{fig:QP_prediction} and \ref{fig:distance} for mass and distance scaling, respectively. In the limit that $T \gg \tau_\mathrm{db}$, we do not expect that the time delay signal should depend much on the integration time and quickly approaches the predicted value after only a few de Broglie times and grows approximately linearly when $T \lesssim \tau_\mathrm{db}$, see Figure \ref{fig:intTime}.
 
The temporal power spectrum of the Shapiro time delay is well described by 
\begin{align}
    P^\mathrm{sh}(f) \propto f^{-8/3} e^{-\tau_\mathrm{db} f / \sqrt{2}} \, .
\end{align}
Figure \ref{fig:PS} shows a comparison to simulated data. Notice that this dependence on frequency is different from the spectrum produced by Doppler shifts \cite{Kim2024}, Compton scale effects \cite{Khmelnitsky:2013lxt}, or stochastic gravitational waves backgrounds from supermassive black hole binaries\cite{Phinney:2001di}. The leading frequency factor comes from the time delay dependence on the size of a spatial mode, where the relevant frequency for a given spatial mode is the inverse crossing time $1/t_{cr}$. The exponential factor comes from the Boltzmann distribution of initial velocities which gives an exponential distribution of initial energy modes. 

This work focuses mainly on the pulsar time delay autocorrelation but a brief qualitative discussion of the spatial correlations is in appendix \ref{appendix:spatial_corr}.

\subsection{Gravitational redshift}
We can likewise approximate the effect of the de Broglie scale effects using the quasiparticle approximation of granular structure. This can be used to approximate the amplitude of the spatial fluctuations of the gravitational potential as 
\begin{align} \label{eqn:phi_rms_approx}
    &\braket{\Phi}_{\mathrm{rms}} / c^2 \approx b \frac{G m_{\text{eff}}}{\lambda_\mathrm{db} \, c^2} \, , \\
    &= b \, G \rho \frac{\hbar^2}{m^2 \sigma^2 \, c^2} \, , \nonumber \\
    &\sim 10^{-20} \left( \frac{10^{-17} \, \mathrm{eV}}{m } \right)^2 \left( \frac{200 \, \mathrm{km/s}}{\sigma} \right)^2 \left( \frac{\rho}{10^7 \, \mathrm{M_\odot / kpc^3}} \right)\,, \nonumber
\end{align}
where $b \approx 2/3$ is an $\sim \mathcal{O}(1)$ constant. The approximation provides an accurate description of the granules in our simulated systems, see Figure \ref{fig:QP_prediction}. In the limit that the pulsar-Earth separation is much greater than the de Broglie wavelength we expect that the $\Phi$ values at each position are uncorrelated, see Figure \ref{fig:distance}. Therefore, the typical redshift, $z$, between Earth and any given pulsar should be of order $z \sim \braket{\Phi}_\mathrm{rms} / c^2$. Now the potential difference between a pulsar and Earth randomly walks about its initial value on the de Broglie timescale. Therefore at some later time we can say 
\begin{align}
    z(t) - z(0) \propto \frac{\braket{\Phi}_\mathrm{rms}} {c^2} \sqrt{\frac{t}{\tau_\mathrm{db}}}
\end{align}

\begin{widetext}
The time delay signal is a cumulative integral of the redshift signal.  The amplitude of the time delay signal should be approximately 
\begin{align}
    \delta t_{rms}^z &\sim \int_0^t z(t) \, dt = \frac{ b\braket{\Phi}_{\mathrm{rms}} \tau_\mathrm{db}}{ c^2} \left( \frac{T}{\tau_\mathrm{db}} \right)^{3/2} \label{eqn:dt_z} \\
    &\sim 4 \times  10^{-4} \left( \frac{10^{-17} \, \mathrm{eV}}{m } \right)^{3/2} \left( \frac{200 \, \mathrm{km/s}}{\sigma} \right)^4 \left( \frac{\rho}{10^7 \, \mathrm{M_\odot / kpc^3}} \right) \, \left( \frac{T}{30 \, \mathrm{yrs}} \right)^{3/2} \mathrm{ns} \nonumber \, .
\end{align}
\end{widetext}
\begin{figure*}[!ht]
	\includegraphics[width = .97\textwidth]{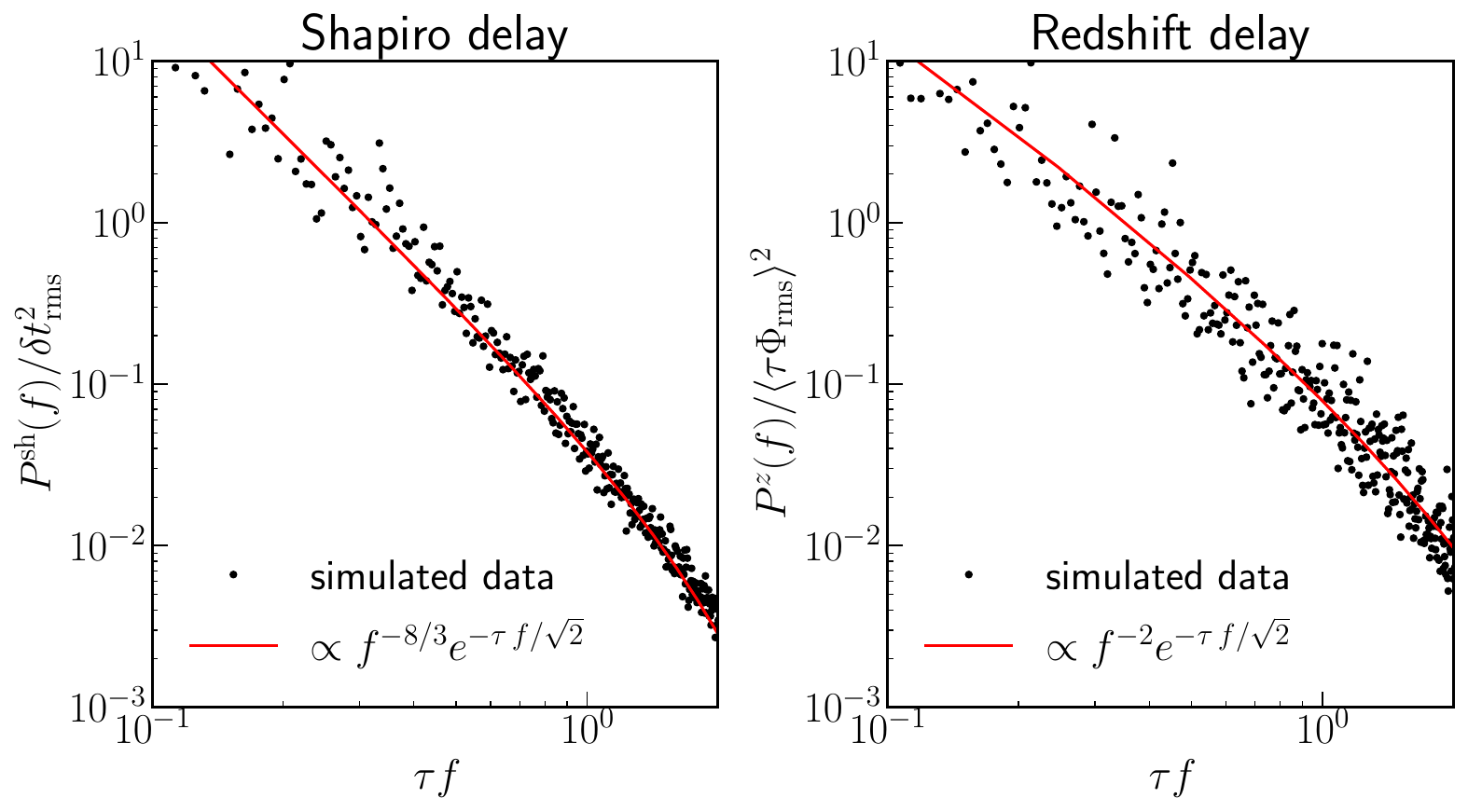}
	\caption{ The temporal power spectrum of the average pulsar Shapiro time delay (right) and average pulsar redshift (left) in a plane-wave box simulation described in section \ref{sec:ICs} with mass $2.5m_{22}$. }
	\label{fig:PS}
\end{figure*} 
The redshift, $z$, observed on Earth for a pulsar at large separations compared to the de Broglie scale, i.e. $D \gg \lambda$, is approximately equal to the $\Phi_{rms}$ times an $\mathcal{O}(1)$ constant. We recover a time dependence similar to the one predicted for the pulsar Doppler shift in \cite{Kim2024}. The time dependence of the signal can be thought of as a result of the integration of random potential fluctuations, see Figure \ref{fig:intTime}.

The temporal power spectrum of the redshift is well described by 
\begin{align}
    P^\mathrm{sh}(f) \propto f^{-2} e^{-\tau_\mathrm{db} f / \sqrt{2}} \, .
\end{align}
Figure \ref{fig:PS} shows a comparison to simulated data. Notice that, as in the Shapiro time delay case, this dependence on frequency is different from the spectrum produced by Doppler shifts \cite{Kim2024}, Compton scale effects \cite{Khmelnitsky:2013lxt}, or gravitational waves.
The leading frequency factor comes from the redshift dependence on the size of a spatial mode, where the relevant frequency for a given spatial mode is the inverse crossing time $1/t_{cr}$. The exponential factor comes from the Boltzmann distribution of initial velocities which gives an exponential distribution of initial energy modes. 

This work focuses mainly on the pulsar time delay autocorrelation but a brief qualitative discussion of the spatial correlations is in appendix \ref{appendix:spatial_corr}.
 
\subsection{Doppler shift}

\begin{figure}[!ht]
	\includegraphics[width = .48\textwidth]{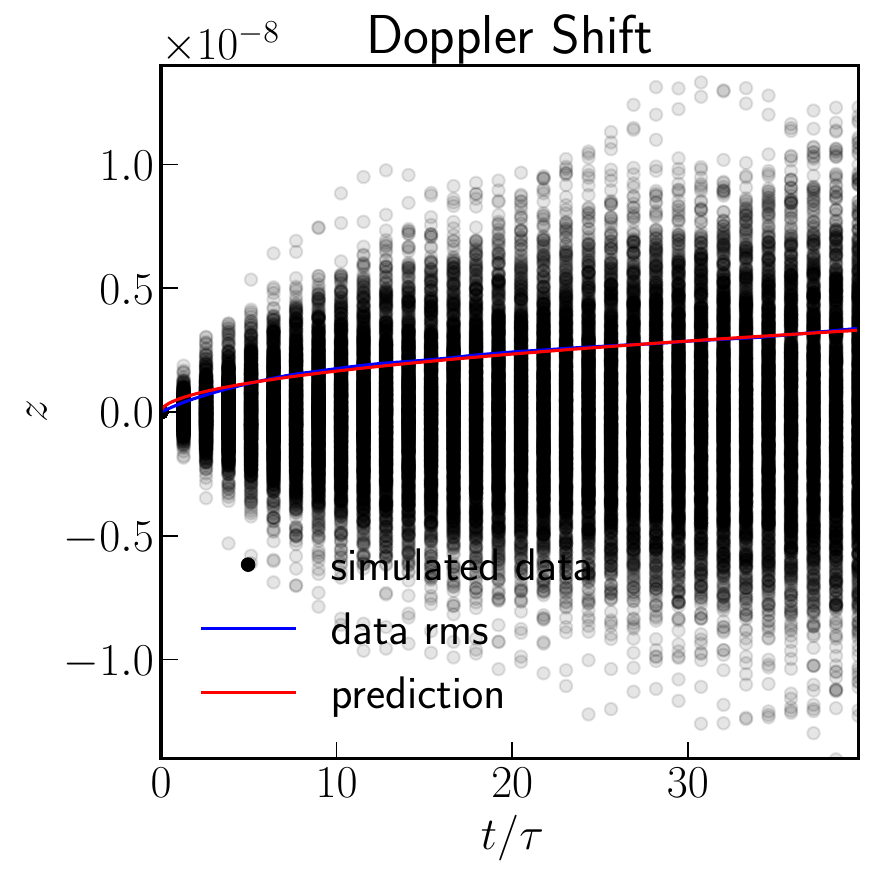}
	\caption{ The redshift due to the Doppler shift of individual simulated pulsars (black dots) is plotted as a function of time. The root-mean-square value of signal is plotted in blue with the prediction (equation \eqref{eqn:doppler_z_over_T}) in red. }
	\label{fig:doppler_vs_time}
\end{figure}

The oscillating de Broglie scale potential provides accelerations to stars in orbit in an ultralight dark matter halo. These accumulated accelerations create a random peculiar velocity between Earth and any given pulsar. This velocity creates a stochastic Doppler shift in the observed pulsar frequency which varies on the de Broglie timescales. 

The spectrum and amplitude of this signal were first worked out in \cite{Kim2024}. They used the quasiparticle approximation to argue that the average acceleration felt by a star in orbit due to a granule should be approximately
\begin{align}
    \bar a \sim \frac{G m_\mathrm{eff}}{\lambda^2} \, .
\end{align}
This creates a Doppler shift over a de Broglie time of $z^D \sim \delta v/c \sim \bar a \tau / c$. These kicks accumulate randomly and result in a total number of kicks $N_\mathrm{enc} / \tau \sim T_\mathrm{obs} / \tau$ where $T_\mathrm{obs}$ is the total observation time. This results in a root-mean-square redshift going as
\begin{align} \label{eqn:doppler_z_over_T}
    z^D_\mathrm{rms}(T) \sim \frac{G m_\mathrm{eff}}{c \lambda \sigma} \sqrt{t / \tau} \, .
\end{align}
We can see in Figure \ref{fig:doppler_vs_time} and Figure \ref{fig:QP_prediction}. that this equation correction predicts the growth of the root-mean-square Doppler shift. The integral of this frequency change then gives an expected time delay signal
\begin{widetext}
\begin{align}
    \delta t^D &\sim (\bar a \tau^2/c) \left( \frac{T_\mathrm{obs}}{\tau} \right)^{3/2} \,   \\
    &= b \frac{G \rho \hbar^{3/2} T^{3/2} }{m^{3/2} \sigma^2 c} \\ 
    &\sim 0.1 \left( \frac{10^{-17} \, \mathrm{eV}}{m} \right)^{3/2} \left( \frac{\rho}{10^7 \, \mathrm{M_\odot / kpc^3}} \right) \, \left( \frac{200 \, \mathrm{km/s}}{\sigma} \right)^2 \, \left( \frac{T}{30 \, \mathrm{yrs}} \right)^{3/2} \mathrm{ns} \, . \nonumber
\end{align}
\end{widetext}
Which is the same result as found in \cite{Kim2024} up a numertion predicts the growth of the root-mean-squareical factor of a few,$b$, which we determine using our simulated data. We note that in \cite{Kim2024} the factors of $\pi$ in the de Broglie wavelength differ from our notation and this also changes the value of the constant $b$.



\section{Observational prospects} \label{sec:observations}
\begin{figure*}[!ht]
	\includegraphics[width = .97\textwidth]{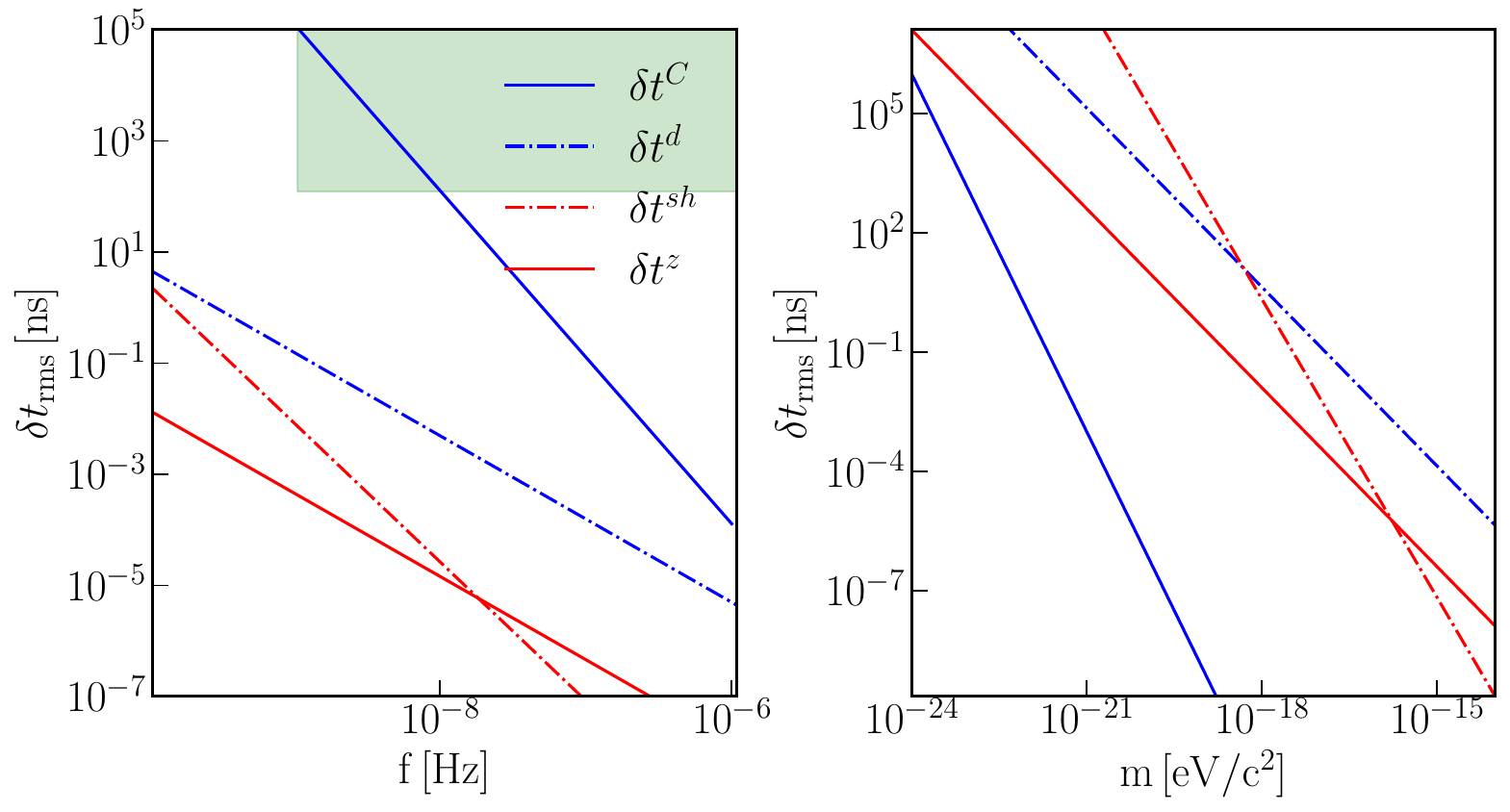}
	\caption{Comparison of the amplitude of the $\delta t_{rms}$ signal as estimated in the de Broglie, Doppler, and Compton cases. Results of previous works, the Doppler \cite{Kim2024} and Compton \cite{Khmelnitsky:2013lxt} cases are shown in blue, and the Shapiro and Redshift time delays as calculated in this work are shown in red. The Compton signal as estimated in equation (3.9) of \cite{Khmelnitsky:2013lxt}, equation \eqref{eqn:dt_c}, is shown in solid blue. The time delay from Doppler shifts as estimated in equation (1) of \cite{Kim2024} is shown in dashed blue. In red we present the Shapiro and redshift time delay, equations \eqref{eqn:dt_sh} and \eqref{eqn:dt_z} respectively. The green area corresponds to the sensitivity of pulsar timing for 100 pulsars are 50 ns precision as estimated in \cite{Khmelnitsky:2013lxt}. The left plot compares the signals as a function of frequency, using the Compton frequency for the Compton signal and de Broglie frequency for the de Broglie and Doppler signals. The right plot shows the amplitude of the signal as a function of the mass. We have assumed that the pulsar-observer separation is $1 \, \mathrm{kpc}$ and that the velocity dispersion of galactic dark matter is $\sigma = 200 \, \mathrm{km/s}$ and the galactic dark matter density is $\rho = 10^7 \, \mathrm{M_\odot / kpc^3}$. In this plot we fix the integration time to $T = 30 \, \mathrm{yrs}$.  }
	\label{fig:dt_c_vs_db}
\end{figure*}
\begin{figure}[!ht]
	\includegraphics[width = .48\textwidth]{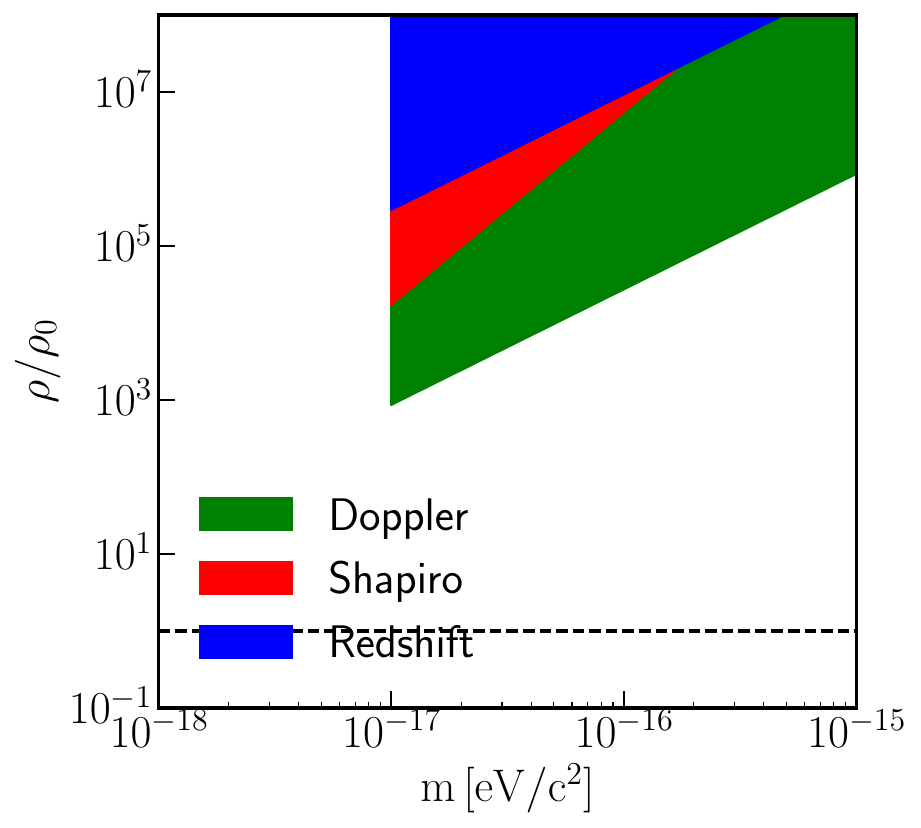}
	\caption{ Approximate exclusion contours for the galactic dark matter density, $\rho$, as a fraction of the approximate measured local dark matter density $\rho_0 = 10^7 \, \mathrm{M_\odot / kpc^3}$. The estimated contours here are approximately similar to other granular constraints, see \cite{Kim_2018,Kim2024}, but far from unity. We have assumed that the pulsar-observer separation is $1 \, \mathrm{kpc}$ and that the velocity dispersion of galactic dark matter is $\sigma = 200 \, \mathrm{km/s}$, the integration time $T = 30 \, \mathrm{yrs}$, and the galactic dark matter density is $\rho = 10^7 \, \mathrm{M_\odot / kpc^3}$, the sensitivity is the same estimate for 100 pulsars with 50 ns precision estimated in \cite{Khmelnitsky:2013lxt}, see also Figure \ref{fig:dt_c_vs_db}. The left edge corresponds to when the de Broglie time is $\tau = T \sim 30 \, \mathrm{yrs}$. }
	\label{fig:contour}
\end{figure}

We can compare the expected amplitude of the signal, $\delta t_\mathrm{rms}$, in the de Broglie and Compton cases. In Figure \ref{fig:dt_c_vs_db}, we compare the time delay signal as a function of mass and frequency for the following effects:
\begin{enumerate}
    \item de Broglie scale Shapiro delay (this work),
    \item de Broglie scale Redshift delay (this work),
    \item de Broglie scale Doppler delay (analytic work in \cite{Kim2024}, simulated in this work),
    \item Compton scale Redshift delay \cite{Khmelnitsky:2013lxt}.
\end{enumerate}
More details on the last effect are in Appendix \ref{sec:appendix_other_pta_work}. To provide an ``apples-to-apples" comparison, we use the same sensitivity curve estimated in \cite{Khmelnitsky:2013lxt}.

We can think of the signal either as a function of mass or as a function of frequency. Figure \ref{fig:dt_c_vs_db} contains both comparisons. If we think of the signals in terms of mass then we expect generally that the de Broglie effects will be larger than the Compton effects at a given mass, and certainly over the mass range where the assumptions we make in this paper are valid, i.e. $m \gtrsim 10^{-22} \, \mathrm{eV}$. This is fundamentally related to the fact that the field is non-relativistic and we expect the portion of the potential that oscillates at the Compton frequency to be suppressed compared to the Newtonian potential by an approximate factor of $\sigma^2/c^2$, see details in \cite{Khmelnitsky:2013lxt}. Therefore, we expect if it is possible to probe de Broglie and Compton effects at a given mass that the de Broglie scale effects will dominate, this is the conclusion of the right panel in Figure \ref{fig:dt_c_vs_db}. 

However, if we consider the signals at a given frequency instead of a given mass we find the opposite. This is because the Compton scale effects oscillate at the Compton frequency, i.e. $f_c= 1/\tau_c =m c^2/ 2 \pi \hbar $, and the de Broglie effects at the de Broglie frequency, $f_\mathrm{db} =1/\tau_{\mathrm{db}}=  m \sigma^2/ 2 \pi \hbar $, which are again related by the same factor of $\sigma^2 / c^2$. Therefore, for a given experiment that may be only sensitive to de Broglie or Compton effects with timescales smaller than the experimental runtime, the Compton scale effects will probe lower masses than the de Broglie ones. And since the signal is typically inversely proportional to the mass this means a relatively smaller signal for the de Broglie scale effect in this case, see the left panel of Figure \ref{fig:dt_c_vs_db}. 

This means that for the de Broglie scale effects studied in this paper, the potential sensitivity of pulsar timing experiments actually benefits more from longer integration times than from increasing the number of pulsars or pulsar timing precision. 

Using the same sensitivity curve estimated in \cite{Khmelnitsky:2013lxt} (which is also the green shaded region in plot \ref{fig:dt_c_vs_db}) we can provide a rough estimate of a constraint on the galactic dark matter density. This is done in Figure \ref{fig:contour}. We find results comparable to other probes of granularity, e.g. \cite{Kim2024, Kim:2024xcr}, but remain far from the observed local dark matter density with current sensitivities. The left edge of the shaded regions in Figures \ref{fig:dt_c_vs_db} and \ref{fig:contour} corresponds to when the relevant timescale (de Broglie time for de Broglie scale effects or Compton time for Compton scale effects) is $30 \, \mathrm{yrs}$. If we look at the slopes of the shaded regions in Figure \ref{fig:dt_c_vs_db} we can again see that increasing the timescales to which we are sensitive quickly increases the constraining power of this method.


\section{Conclusions} \label{sec:conclusions}

In this paper, we have studied the time delay in pulsar signals created by the de Broglie granular structure of ultralight dark matter. We have studied this system semi-analytically using the quasi-particle approximation and compared predictions against simulations of mock pulsars in oscillating granular density fields. We have calculated the amplitude of the expected root-mean-square Shapiro time delay, gravitational redshift time delay, and Doppler shift time delay which are given as
\begin{widetext}
\begin{align}
    \delta t_\mathrm{rms}^\mathrm{sh} &\sim 7 \times 10^{-3} \left( \frac{10^{-17} \, \mathrm{eV}}{m} \right)^{1/2} \left( \frac{200 \, \mathrm{km/s}}{\sigma} \right)^{5/2} \left( \frac{\rho}{10^7 \, \mathrm{M_\odot / kpc^3}} \right) \left( \frac{D}{\mathrm{kpc}} \right)^{1/2} \, \mathrm{ns} \, , \nonumber \\
    \delta t_\mathrm{rms}^\mathrm{z} &\sim 4 \times  10^{-4} \left( \frac{10^{-17} \, \mathrm{eV}}{m} \right)^{3/2} \left( \frac{200 \, \mathrm{km/s}}{\sigma} \right)^4 \left( \frac{\rho}{10^7 \, \mathrm{M_\odot / kpc^3}} \right) \, \left( \frac{T}{30 \, \mathrm{yrs}} \right)^{3/2} \mathrm{ns}\,, \nonumber \\
    \delta t_\mathrm{rms}^\mathrm{D} &\sim 0.1 \left( \frac{10^{-17} \, \mathrm{eV}}{m} \right)^{3/2} \left( \frac{200 \, \mathrm{km/s}}{\sigma} \right)^2 \left( \frac{\rho}{10^7 \, \mathrm{M_\odot / kpc^3}} \right) \, \left( \frac{T}{30 \, \mathrm{yrs}} \right)^{3/2} \mathrm{ns}\,, \nonumber
\end{align}
\end{widetext}
respectively. Current pulsar timing experiment sensitivities are unlikely to be able to detect the time delays, see Figures \ref{fig:dt_c_vs_db} and \ref{fig:contour}. However, de Broglie scale effects on observable timescales could potentially provide sensitivity to mass scales higher than those probed thus far by small-scale structure. The current largest limitation appears to be the longest de Broglie time to which experiments are sensitive, presumably the de Broglie time which is approximately their own runtime. Experiments with longer runtimes or somehow sensitive to longer timescales would substantially improve the signal from the effects studied here. Therefore, this remains an interesting signal for possible future experiments. 

We note that the time delay signal has a unique temporal power spectrum depending on the velocity dispersion of the galactic dark matter, see Figure \ref{fig:PS} and a unique dependence on observer-pulsar separation, see Figure \ref{fig:distance}. This means that the detection of such a signal would provide strong evidence for the existence of ultralight dark matter that would be easily distinguished from other potential sources. We also note that the signal is most prominent in the mass range of $\gtrsim 10^{-17} \, \mathrm{eV}$, which are much higher than it is currently possible to probe with small-scale structure arguments alone. 

Finally, we note a few potential caveats of this work and its sensitivity to extensions of the vanilla one classical uniform ultralight dark matter field model. All time delays are proportional to the amplitude of density fluctuations, they are sensitive to extensions of ultralight dark matter which affect the amplitude of these fluctuations, e.g. multiple fields \cite{Gosenca2023}, higher spins \cite{Amin2022}, and quantum corrections \cite{Eberhardt2023}. Additionally, here we have assumed that the ultralight dark matter is approximately uniformly distributed between us and the pulsars. If additional substructure exists below that scale, for example in mini-clusters \cite{Ellis_2022}, then the approximations made in this work would likely not be accurate and a more detailed analysis would be required. Likewise, any dark matter substructure can produce similar time delays, see for example \cite{Schutz_2017}. However, we would expect that the timescales and temporal power spectrum to be different than the ultralight dark matter case. We leave this as potentially interesting future work. 

As pulsar timing arrays and other experiments sensitive to dark matter metric perturbations continue to improve the effect of granule oscillations on observable timescales will remain an interesting future direction of study.

\section*{Acknowledgments}

We would like to thank Mustafa Amin, Gabi Sato-Polito, and Hyungjin Kim for useful discussions. This work is supported by World Premier International Research Center
Initiative (WPI Initiative), MEXT, Japan.

\appendix

\section{Estimating root-mean-square Shapiro delay} \label{app:approxShapiro}

The root-mean-square Shapiro delay in the weak gravity limit of a signal sent from $r_e$ to $r_p$ is given as a line integral of the potential along the unperturbed path $x(t)$. 
\begin{align}
    \delta t = \int_{r_e}^{r_p} \frac{dx(t)}{c} \frac{\delta \Phi(x(t), t)}{c^2} \, .
\end{align}
Where we will only be concerned with the potential fluctuations, $\delta \Phi(r,t) = \Phi(r,t) - \Phi_\mathrm{sm}$, around the smooth potential $\Phi_\mathrm{sm}$. Let us say that these potential fluctuations have root-mean-square amplitude, $\Phi_\mathrm{rms}$. We also know that the correlations between fluctuations go to $0$ on spatial scales large compared to the de Broglie wavelength, $\lambda_\mathrm{db}$, and temporal scales large compared to the de Broglie time, $\tau_\mathrm{db}$ and therefore we can say
\begin{align} 
    \braket{\delta \Phi(r) \, \delta\Phi(r + \Delta r)}  \left\{
\begin{array}{ll}
      \ll \delta \Phi^2   \,,& \Delta r \gg \lambda_\mathrm{db} \\
      \sim \delta \Phi^2 \,,& \mathrm{else}  \\
\end{array} 
\right. \, . \label{eqn:correlations_Phi}
\end{align} 
Note a similar equation exists at a given position in time with $r \rightarrow t$ and $\Delta r/c \gg \tau_\mathrm{db}$. Therefore, for a path with total length $L \gg \lambda$ and total light travel time $T \gg \tau_\mathrm{db}$ we can write the variance of the temporal fluctuations as
\begin{widetext}
\begin{align}
    \braket{\delta t}_\mathrm{rms}^2 &= \int_{r_e}^{r_p}  \frac{dx(t)}{c}  \int \frac{dx'}{c}  \frac{\braket{ \delta \Phi(x(t), t) \, \delta \Phi(x(t) + x', t + x'/c) }}{c^4} \nonumber  \\
    &\approx \sum_{j = 1}^{L/\lambda_\mathrm{db}} \frac{\lambda^2}{c^2} \frac{\delta\Phi^2_\mathrm{rms}(r_e + j \lambda_\mathrm{db}, \, j \lambda_\mathrm{db} / c)}{c^4} \label{eqn:dt_rms_1} \, .
\end{align}
\end{widetext}
Where the travel time is $\Delta t = j \lambda_\mathrm{db} / c$ to position along the path $x(j \lambda_\mathrm{db} / c) = j \lambda_\mathrm{db}$. Now if the potential is in a steady-state (i.e. the statistics of the potential fluctuations are not functions of time) and spatially constant as we expect to be true on the time and length scale of our observations of pulsars, then we can say that 
\begin{align}
    \braket{ \Phi(r + \Delta r,t + \Delta t) }_\mathrm{rms} = \braket{ \Phi(r,t) }_\mathrm{rms} \label{eqn:dt_rms_2} \, .
\end{align}
For $\Delta t \gg \tau_\mathrm{db}$. So in the limit that $L / c \gg \tau_\mathrm{db}$ and $L \gg \lambda_\mathrm{db}$ we can approximate equation \eqref{eqn:dt_rms_1} as
\begin{align}
    \braket{\delta t}_\mathrm{rms} \approx \frac{\delta \Phi_\mathrm{rms}}{c^2} \frac{\lambda_\mathrm{db}}{c} \sqrt{L / \lambda_\mathrm{db}} \, ,
\end{align}
which involves applying both the steady-state in time and spatially homogeneous approximations in equation \eqref{eqn:dt_rms_2}. 

When performing the simulations however we do not need to make the spatially homogeneous approximation. Instead we can simply say that $\braket{ \Phi(r,t + \Delta t) }_\mathrm{rms} = \braket{ \Phi(r,t) }_\mathrm{rms}$, and in this case write
\begin{widetext}
\begin{align}
    \braket{\delta t}_\mathrm{rms}^2 &= \int_{r_e}^{r_p}  \frac{dx(t)}{c}  \int \frac{dx'}{c}  \frac{\braket{ \delta \Phi(x(t), t) \, \delta \Phi(x(t) + x', t + x'/c) }}{c^4} \nonumber  \\
    &\approx \int_{r_e}^{r_p}  \frac{dx(t)}{c^2}   \frac{\delta \Phi_\mathrm{rms}^2(x(t), t)}{c^4}\,  \\ 
    &\approx \int_{r_e}^{r_p}  \frac{dx}{c^2}   \frac{\delta \Phi_\mathrm{rms}^2(x)}{c^4} \, \\
    \braket{\delta t}_\mathrm{rms} &= \int_{r_e}^{r_p}  \frac{dx}{c^2}   \frac{\delta \Phi_\mathrm{rms}}{c^2} \,
\end{align}
\end{widetext}
Where in the first line of the above equation we have used equation \eqref{eqn:correlations_Phi} and in the second line we have used the temporal portion of equation \eqref{eqn:dt_rms_2}.

\section{Density power spectrum}
\begin{figure}[!ht]
	\includegraphics[width = .48\textwidth]{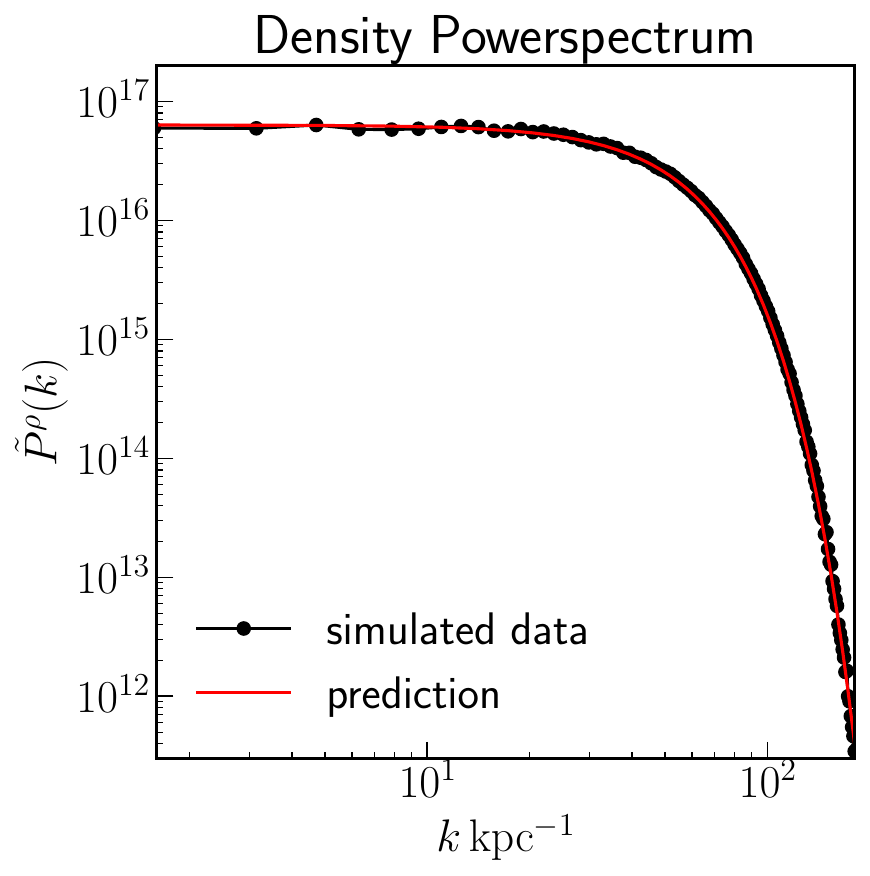}
	\caption{ The power spectrum of the density for simulated data and comparison with equation \eqref{eqn:rhoPS}. We can see that this prediction clearly predicts the correct shape of the spectrum. The normalization of the power spectrum plotted is arbitrary. The field simulated had a mass $m = 2.5 \times 10^{-22} \,\mathrm{eV}$. }
	\label{fig:rho_PS}
\end{figure}
\label{sec:appendix_density_ps}
The power spectrum of the dark matter density is an important quantity to model correctly when approximating systems without running full-scale cosmological or isolated halo simulations. Previous work has demonstrated the ability of superimposed plane waves to approximate the statistics of overdensities around the radial density profile \cite{Dalal_2021}. 

Here we model the density as a superposition of plane waves drawn from a Maxwell-Bolzmann distribution with the approximate local dark matter dispersion, $\sigma \sim 200 \, \mathrm{km/s}$. This means that the power spectrum of the density can be easily estimated as 
\begin{align}
    P^\rho (k) \propto e^{-\hbar^2 k^2 / 2\sqrt{2} \sigma^2 m^2} \label{eqn:rhoPS}
\end{align}
Where the extra $\sqrt{2}$ in the argument of the exponential comes from the convolution in $\rho \propto |\psi|^2$. We have used the fact that $v = \hbar k / m$. Note that the power spectrum in our simulations is consistent with previous similar work \cite{Dalal_2021}, see Figure \ref{fig:rho_PS}.

\section{Spatial correlations} \label{appendix:spatial_corr}
  
\begin{figure*}[!ht]
	\includegraphics[width = .97\textwidth]{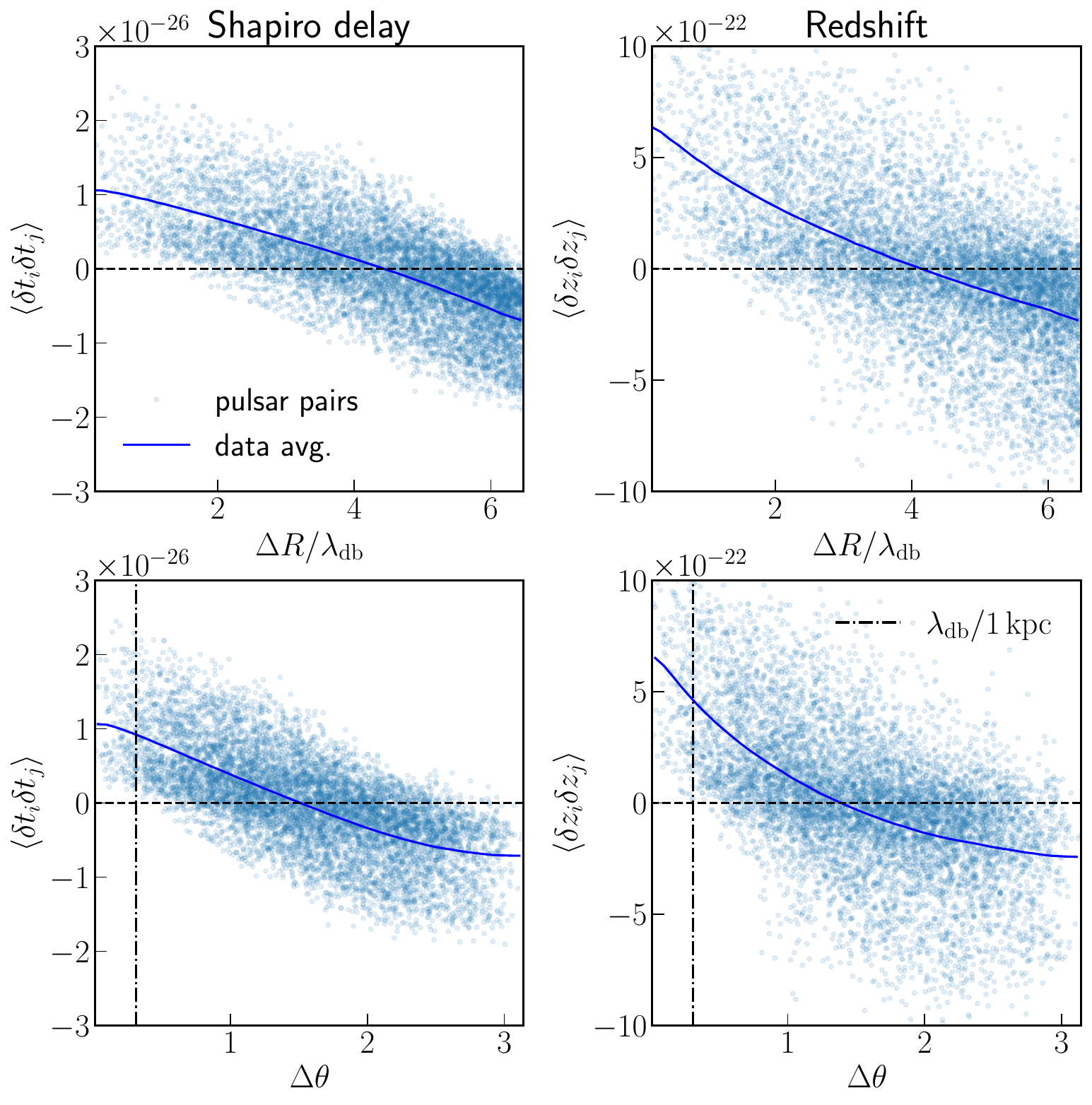}
	\caption{ Here we plot the time averaged spatial and angular correlation function for a simulation where all the pulsars are fixed at a distance of $D = 1 \, \mathrm{kpc}$ from the observer. The left column contains the correlations in the Shapiro time delay and the right column the correlations in the redshift. The first row is spatial correlations and the second angular. We see that the correlation decays on approximately the de Broglie scale, $\lambda_\mathrm{db}$. In this simulation $m = 1.95 \times 10^{-22} \, \mathrm{eV}$ and the total integration time is $T = 60 \, \tau_\mathrm{db}$. }
	\label{fig:corr}
\end{figure*}
In this appendix, we discuss the spatial correlations in the system qualitatively. 
We plot the time averaged spatial and angular correlations of a simulation in Figure \ref{fig:corr}. We can see that the correlation has the expected decay on the de Broglie scale. 

\section{Other ULDM effects on pulsar timing}
\label{sec:appendix_other_pta_work}

In this section, we summarize the effects already described in the literature in the context of this study, namely the Compton scale redshift and the Doppler shift.

\subsection{Compton scale redshift}

The Compton scale oscillations of an ultralight dark matter field are known to create an oscillating potential which can be approximated as
\begin{align}
    \Phi(\vec x,t) = \Phi_0(\vec x,t) + \pi \frac{G \rho}{m^2} \cos{(\omega_c t + 2 \alpha(\vec x))} \,,
\end{align}
where $\Phi_0$ solves the non-relativistic Poisson equation and the second term represents a component of the potential which varies on the Compton timescale, with $\omega_c$ the Compton frequency. This creates a time varying potential difference between Earth and a pulsar which then creates a gravitational redshift in the observed pulsar frequency. The amplitude of this signal was first worked out in \cite{Khmelnitsky:2013lxt}.

The Compton scale portion of the potential is suppressed by a factor $\sim v^2/c^2$ but was estimated to have an observable effect on pulsar timing for masses $\sim 10^{-23} \, \mathrm{eV}$ by \cite{Khmelnitsky:2013lxt,Porayko:2018sfa}.

Using equations (3.8) and (3.9) in \cite{Khmelnitsky:2013lxt}, the amplitude of the Compton signal is about 
\begin{align}
    \delta t_{rms}^C = 10^{3} \left( \frac{10^{-23} \, \mathrm{eV}}{m} \right)^{3} \left( \frac{\rho}{0.3 \, \mathrm{GeV / cm^3}} \right) \, \mathrm{ns} \, . \label{eqn:dt_c}
\end{align}

\bibliography{BIB}

\end{document}